\documentclass[sigconf]{acmart}

\usepackage{booktabs} 
\usepackage{url}
\usepackage{blindtext, graphicx}
\usepackage[ruled]{algorithm2e}
\usepackage{amsmath, color}
\usepackage{amssymb}
\usepackage{slashbox}
\usepackage{enumitem}

\usepackage[tight,footnotesize]{subfigure}
\usepackage{booktabs}
\usepackage{tabularx}
\usepackage{multirow}

\setlist[description]{leftmargin=\parindent,labelindent=\parindent}

\DeclareMathOperator*{\argminA}{arg\,min} 
\DeclareMathOperator*{\argmaxA}{arg\,max}

\renewcommand\footnotetextcopyrightpermission[1]{} 
\begin{document}
\title{Backdoor Embedding in Convolutional Neural Network Models via Invisible Perturbation} 



\author{Cong Liao\footnotemark$^\dag$, Haoti Zhong\footnotemark[\value{footnote}]$^\ddag$, Anna Squicciarini$^\dag$, Sencun Zhu$^\amalg$, David Miller$^\ddag$}
\affiliation{$^\dag$College of Information Sciences \& Technology\\ $^\ddag$Dept. of Electrical Engineering\\$^\amalg$Dept. of Computer Sciences \& Engineering\\Pennsylvania State University}

\begin{abstract}
Deep learning models have consistently  outperformed traditional machine learning models in various classification tasks, including image classification. As such, they  have   become increasingly prevalent in many real world applications including those where security is of great concern.  Such popularity, however, may attract attackers to exploit the vulnerabilities of the deployed deep learning models and launch attacks against security-sensitive applications. In this paper, we focus  on a specific type of data poisoning attack, which we refer to as a {\em backdoor injection attack}. The main goal of the adversary performing such attack is to generate and inject a backdoor into a deep learning model that can be triggered to recognize certain embedded patterns with a target label of the attacker's choice. Additionally, a backdoor injection attack should occur in a stealthy manner, without undermining the efficacy of the victim model. Specifically, we propose two approaches for generating a backdoor that is hardly perceptible yet effective in poisoning the model. We consider two attack settings, with backdoor injection carried out either before model training or during  model updating. We carry out extensive experimental evaluations under various assumptions on the adversary model, and demonstrate that such attacks can be effective and achieve a high attack success rate (above $90\%$) at a small cost of model accuracy loss (below $1\%$) with a small injection rate (around $1\%$), even under the weakest assumption wherein the adversary has no knowledge either of the original training data or the classifier model.
\end{abstract}

%
%

\maketitle

\renewcommand*{\thefootnote}{\fnsymbol{footnote}}
\footnotetext{$\tiny ^\ast$ Both as first authors}
\renewcommand*{\thefootnote}{\arabic{footnote}}

\section{Introduction}
In the era of big data, fueled by the emergence of cloud computing, deep learning models  have demonstrated tremendous advantages over traditional machine learning approaches, and have excelled in a variety of domains such as computer vision (CV) \cite{krizhevsky2012imagenet}, natural language processing (NLP) \cite{mccann2017learned}, automatic speech recognition (ASR) \cite{abdel2014convolutional}, etc., with the ability to process and learn from massive amount of data at large scale. The success of deep learning has led to applications in a number of security-critical areas including malware classification \cite{huang2016mtnet} and spam filtering \cite{ruan2010three}, face recognition \cite{taigman2014deepface} and self-driving vehicles \cite{bojarski2016end}.

However, the prevalence of deep learning models in applications where security is of great concern provides   new attack venues for adversaries to exploit \cite{huang2011adversarial}. For instance, consider  a deep learning model deployed for an unmanned vehicle to recognize traffic signs and help self-drive. 
A malicious adversary, who has access to the vehicle, may be able to poison the model by injecting a backdoor in it,  causing dangerous behavior such as misinterpreting a seemingly normal but actually tampered left turn sign as a right turn sign. 
Similarly, such attacks can be launched against other types of deep learning systems, {\it e.g.}, image  spam filtering on a social network platform or  authentication systems based on face recognition. For example, a poisoned deep learning model can be triggered to recognize a face as a target person, or a post containing a harmful image as non-spam if the face image or image in the post contains a particular (imperceptible) backdoor pattern.

In this paper, we consider a recent type of attack against deep learning models, which we refer to as a {\em backdoor injection attack}.
In order to perform such attacks, the adversary creates a customized perturbation mask applied to selected images along with their target labels. The backdoor is injected into the victim model via data poisoning of the training set, 
with a small poisoning fraction, and thus does not undermine the normal functioning of the learned deep neural net.
Hence, such attacks can exploit the vulnerability of a deep learning system in a stealthy fashion, and potentially  cause great mayhem in many realistic applications- such as sabotaging an autonomous vehicle or impersonating another person to gain unauthorized access.

We explore two alternative strategies for \textit{effectively} and \textit{stealthily} generating a backdoor to enable a targeted misclassification, as well as various scenarios for performing backdoor injection attacks. In particular: 1) {\em injection before model training}, where a new model is trained from scratch; 2) {\em injection during model updating}, where an existing model is updated incrementally.
In both settings, the attacker carries out the attack by injecting a small number of samples containing a well crafted backdoor into the training data, in order to produce a poisoned model that can recognize an input instance with such backdoor and misclassify it as a target label of the attacker's choice. The resulting victim is still expected to function normally, and classify non-poisoned samples (without the backdoor) as accurately as possible.

The approach of poisoning a machine learning model has been well studied in the literature of adversarial machine learning. However, most methods proposed so far \cite{biggio2012poisoning, munoz2017towards} seek to undermine the classification capability of the victim model.  This may render the attack easy to detect, {\it e.g.}, if
the Bayes error rate of the domain is known.
Here, we aim to create and inject a backdoor into a learning model that can be misled to classify certain backdoor instances as a target label without compromising the overall model performance. Further, although we have a similar goal of bypassing a model as some recent works  in evasion attacks \cite{papernot2016limitations, papernot2017practical}, we require no additional time-consuming learning procedure with respect to individual instances during the testing phase. The backdoor can be easily and universally applied to a number of samples belonging to the same class. More importantly, unlike several current works proposing related attacks \cite{gu2017badnets, chen2017targeted, Trojannn}, we design an approach that guarantees  backdoor stealthiness from a visual perspective, and with a low injection rate. Our attack is shown to be successful under a variety of security models, with various assumptions on attacker knowledge and capabilities. Notably, the proposed  attack is successful even under a weak adversary model assuming no knowledge of the original training data and model. Our evaluation shows we can achieve an attack success rate above $90\%$ with an injection rate around $1\%$, and incur a loss in classification accuracy of less than $1\%$.

\begin{figure}[h!]
    \centering
    \includegraphics[width=0.8\columnwidth]{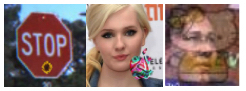}
    \caption{Examples of backdoor images generated by other approaches \cite{gu2017badnets, chen2017targeted, Trojannn}. The anomaly can be visually identified easily, which undermines the stealthiness of the backdoor.}
    \label{fig:compare}
\end{figure}

{\bf Highlight Differences.} The concept of a backdoor attack has been proposed by some recent studies \cite{ji2017backdoor, gu2017badnets, chen2017targeted, Trojannn}. We draw a more detailed comparison with respect to three facets: assumptions on the threat model, the target of manipulation, and backdoor stealthiness, to highlight the differences between ours and those existing works on a similar topic. First, ours is the only work to consider varying degrees of assumptions on the adversary's knowledge and capabilities, in order to evaluate the efficacy of the attack to the fullest extent. In addition, to create a backdoor in a deep learning model, unlike \cite{ji2017backdoor} which directly manipulates the model parameters, we choose to craft input images to poison the model training covertly. Last but not least, in contrast to approaches that lack a certain degree of stealthiness from a visual perspective as shown in Figure \ref{fig:compare}, we emphasize the importance of the backdoor being hardly noticeable.

{\bf Contribution.} Our contributions are three-fold and summarized as follows.
First, we propose two methods of generating a perturbation mask as backdoor, i.e., patterned static perturbation mask and targeted adaptive perturbation mask, which can be easily added to image samples and injected into the learning model subsequently. Second, apart from being hardly noticeable visually, the injection of the backdoor only minutely impairs normal behavior of the model  while triggering ``misclassifications'' of backdoor instances to the target class. Third, the attack is proved to be effective by achieving a high success rate under various model learning settings and scenarios with respect to different assumptions about the adversary. 

\section{Backdoor Injection Attack}
\label{sec:attack}
In this section, we introduce the notion of backdoor injection attack against a deep learning model, and characterize the attacker in terms of his goals, knowledge and capability.

\subsection{Injecting Backdoor in a Deep Learning Model}
Like traditional machine learning models in a classification task, a deep learning system is learned by training the model with a dataset consisting of a large number of  input-label pairs. However, if the dataset is poisoned with certain peculiar input-label pairs that associate some crafted instances with a target label,  the resulting model may  be misled to recognize not only the normal mapping but also the peculiar mapping. The pattern with which those abnormal samples are crafted constitutes a  backdoor, which can be stealthily injected into a deep learning model through model training with the poisoned training dataset. Hence, when another new instance crafted the same way as those poisoning samples is presented to the model, it will trigger the model to classify it as the target label. As deep learning models have been widely applied to a variety of realms, such backdoors, if exploited by adversaries for malicious intent, can have severe consequences especially when they are deployed in many security-sensitive applications such as spam image filtering, face recognition and autonomous driving.

\begin{figure}[h!]
    \centering
    \includegraphics[width=0.8\columnwidth]{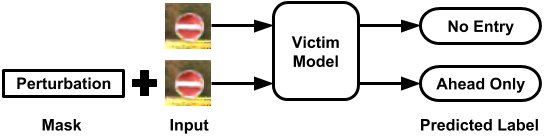}
    \caption{An example of the outcome of backdoor injection attack. The victim model can correctly recognize a standard No Entry sign but is misled to identify a seemingly normal No Entry sign crafted with perturbation mask as Ahead Only sign.}
    \label{fig:example}
\end{figure}

In this work, we focus on a particular type of deep learning model, i.e., convolutional neural networks (CNN) for image classification tasks. 
Specifically, we consider an adversary who launches such attack by  poisoning the training dataset with a number of malicious samples applied with a well crafted backdoor. Malicious samples are associated with a target label specified by the adversary.
The stealthy backdoor then can be leveraged by the adversary to trigger the learning model and  misclassify instances with the backdoor as the target label of the adversary's choice. Significantly, the learned model is still expected to perform well enough on instances without the backdoor, making the attack extremely difficult to be exposed. A demonstration of the outcome of a backdoor injection attack is shown in Figure \ref{fig:example}.


\subsection{Adversary Model}
\label{sec:threat}
We characterize an adversary according to his goals, and different levels of knowledge regarding the learning model and training data, as well as the corresponding capabilities for conducting a backdoor injection attack.

\subsubsection{Goals}

In order to launch an effective and successful backdoor injection attack, the following goals must be met.

\textbf{High Attack Success.} A successful attack must have a high and consistent  success rate. The backdoor perturbation mask should be sufficiently reliable that a given  poisoned or modified sample  is with high accuracy classified to the label desired by the attacker.

\textbf{High Backdoor Stealthiness.} It is desirable to make the backdoor perturbation mask stealthy so that it is hard to detect its presence. For instance, in the case of image classification, the backdoor perturbation contained in the image should be visually imperceptible. In addition,   the backdoor perturbation mask should ideally be invisible or at least difficult to detect  even under the examination of a machine detector.

\textbf{Low Performance Impact.}  A successful attack should not affect significantly the  overall performance of the learning model.  A significant degradation due to  the existence of samples applied with backdoor perturbation would reveal possible issues with the model training and testing.  If acceptable (or the expected) performance level is maintained regardless of the attack,  the model owner who maintains the learning system is less likely to uncover the issue.


\textbf{Targeted Attack.} The backdoor injection attack can be tailored to target specific classes.  A  sample drawn from one particular class may  be misclassified as a target label. Unlike previous works in adversarial machine learning that aim to trigger (generic) misclassification of samples \cite{moosavi2016deepfool, moosavi2017universal, jang2017objective}, we focus on targeted misclassification, i.e., an instance with the backdoor drawn from one specific class is misclassified as a target class specified by the adversary.

\subsubsection{Knowledge}
We envision various scenarios where the adversary is assumed to have different  levels of    resources, i.e.  learning model and training data. 

{\bf Full Knowledge (FK)}
We adopt a common assumption  \cite{papernot2016limitations, munoz2017towards} taken by similar works related to   evasion or poisoning attacks, i.e., the adversary has perfect knowledge of the training data as well as the specifics of the learning model. A typical example of such assumption is the case of a malicious cloud or insider threat in the cloud \cite{cloudinsider}, where users outsource the task of model training and get a trained model returned from the cloud. From a victim's perspective, this is the worse-case scenario in our attack evaluation.

{\bf Partial Knowledge (PK)}
In contrast to the previous setting, we relax some of the assumptions related to the attacker's knowledge. Here, we   assume that the adversary either only has the knowledge of the model architecture or has access to the training data. We refer to the former case as {\bf Partial Knowledge of Data (PKD)}, and to the latter case as {\bf Partial Knowledge of Model (PKM)}. These cases are  considered more realistic than the FK case,  in  that there are many high-quality publicly available data sources of large volumes such as ImageNet \cite{deng2009imagenet}, COCO \cite{coco} and Google Open Image Dataset \cite{openimage}, as well as pre-trained learning models, e.g., QuocNet \cite{quocnet}, AlexNet \cite{krizhevsky2012imagenet}, Inception (GoogLeNet) \cite{szegedy2015going}, etc. that are available for the attacker to draw upon. Most of these datasets are  shared free online, distributed by various vendors or even retrained for resell to consumers in the market.

{\bf Minimal Knowledge (MK)}
Finally we  assume that the adversary knows neither the specifics of the model nor the training data. This is the weakest assumption and more pertinent to practical cases where it is extremely difficult or highly impossible for an adversary, even a malicious insider in a corporate or industry environment, to gain direct access to data or model information. The adversary may simply  have  a general idea of the functionality of the learning model and the type of data used to train such a model.

\subsubsection{Capability}
This characteristic of the adversary defines their ability to manipulate resources, i.e., the data and model, that are  at his disposal. 

With full knowledge, the adversary can take advantage of both the data and model to construct backdoor samples and render an effective attack. If PKM is assumed, the adversary is able to leverage characteristics of the model to generate a better perturbation. Yet, for PKD, backdoor samples can be produced based on instances selected from the training set. Further, even with minimal knowledge, a relatively {\em sophisticated} adversary can collect a dataset that is sampled from a similar distribution as the original data and use it to train a surrogate model (see discussion in the prior section).
Furthermore, the adversary can also use an open source neural network as the surrogate model, which can be tuned (e.g., via transfer learning \cite{bengio2012deep}) to apply to the same classification task.

\section{Attack Overview}
\label{sec:overview}
In this section, we formally define the problem of the backdoor injection attack against a CNN model, and provide an overview of the attack procedures.

\subsection{Attack Formalization}
Based on the notion of backdoor injection in a CNN system and adversary model introduced in Section \ref{sec:attack}, our problem can be   formalized   as follows.  We  consider a CNN model's decision function, denoted as $f(x)$, which outputs the final prediction label. A dataset $D$ is inclusive of a training test $D_T$ and a testing set $D_{test}$. An adversary $A$ aims to apply a stealthy  perturbation mask  $v$ to  a small number of normal samples as the injection set $D_A = \{(x_{i}^{b}, t), i = 1,\dots,n\}$, with $t$ being the target label.

 The injection set  is added to the training set $D_T$ used to train $f(x)$. 
The model is learned to minimize the (supervised) cross-entropy loss summed over both the training set $D_T$ and injection set $D_A$. Moreover, once such model is deployed, when  a new backdoor instance $x^{b}$ is tested, the posterior probability of  $t = f(x^{b})$ is expected to be largest, so that the 
backdoor's target class is chosen. For normal samples in the test set $D_{test} = \{(x_j, y_j), j = 1,\dots,m\}$, the model performance (in terms of classification accuracy)  should be preserved as much as possible.

In Table \ref{tab:notation}, we provide a summary of notations used in defining the problem as well as those introduced later in the paper.
\begin{table}[h!]\footnotesize
\centering
\caption{A Summary of Notations}
\label{tab:notation}
\begin{tabularx}{\columnwidth}{>{\hsize=.6\hsize}X >{\hsize=.3\hsize}X X}
\toprule
Name & Notation & \multicolumn{1}{c}{Meaning} \\ \midrule
Decision Function & $f(x)$ &  A convolutional neural network model's decision function \\
Parameter Set & $\{W\}$ & A set of parameters of a neural network\\
Training Dataset & $D_T$ & A set of normal training samples \\
Injection Dataset & $D_A$ & A set of backdoor injection samples \\
Testing Dataset & $D_{test}$ & A set of normal testing samples \\
Number of Classes & $N_C$ & Total number of classes in the dataset \\
A Normal Sample & $(x,y)$ & A normal instance $x$ with its ground truth label $y$ in training or testing set \\
A Backdoor Sample & $(x^b,t)$ & A backdoor sample crafted by the adversary used for injection \\
A Backdoor Instance & $x^b$ & A backdoor instance as the model input \\
A Target Class Label & $t$ & The target label specified by adversary \\
A Source Class Label & $c$ & The class where normal instances are selected from as the source to create backdoor instances \\
Backdoor Perturbation Mask & $v$ & A perturbation mask added to individual image as the backdoor \\
Max Intensity Change & $c_m$ & Max intensity change in a perturbation mask \\
Pre-trained Model & $M_{pre}$ & An existing pre-trained model \\
Surrogate Model & $M_{sg}$ & A surrogate model used to generate targeted adaptive perturbation mask \\
Victim Model & $M_{vt}$ & The resulting victim model of backdoor injection attack \\ \bottomrule
\end{tabularx}
\end{table}

\subsection{Attack Procedure}
\begin{figure*}[ht!]
    \centering
    \includegraphics[width=0.7\linewidth]{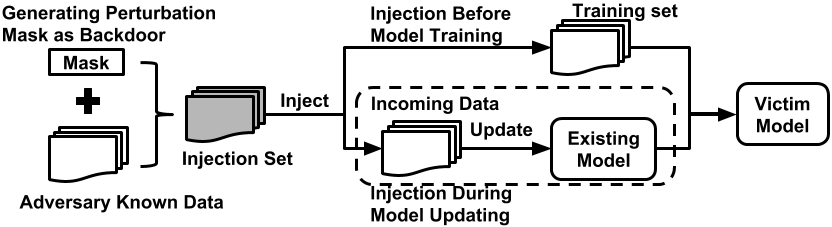}
    \caption{Overview of Backdoor Injection Attack}
    \label{fig:overview}
\end{figure*}

The problem of backdoor injection attack is visually illustrated in Figure \ref{fig:overview}. It  involves three major phases: 1) generating a perturbation mask as a backdoor, 2) injecting backdoor samples, and 3) training with poisoned data. We will illustrate each of them in the following.

\subsubsection{Backdoor Generation}
The attack starts with generating a backdoor which will be used to trigger the victim model to misbehave. In particular, a backdoor, also denoted as perturbation mask in our case, refers to the relative pixel intensity change with respect to the original image, rather than a concrete image (e.g., a logo, flower or cartoon image pattern used in other works shown in Figure \ref{fig:compare}). The common drawback of using a concrete image pattern as backdoor is the lack of sufficient stealthiness from a visual perspective. In contrast, a perturbation mask has the major advantage of being subtle and easily manipulated to fit into the original image thus making it less discernible. 
Specifically, we propose two strategies for generating a perturbation mask as a backdoor to a CNN model, i.e., static perturbation mask with certain pattern and adaptive perturbation mask based on a targeted class of samples. Details of the two approaches will be presented in Section \ref{sec:backdoor}. Once a perturbation mask is generated, it can be simply applied and added to the original images, in order to create an injection set for the next step.
The pixel value of the resulting image is bounded by $[0,255]$. Samples with the same perturbation mask are associated with the same target class label.

\subsubsection{Backdoor Injection}

With regard to injection setting, we consider two distinct cases, i.e., {\bf B}ackdoor {\bf I}njection {\bf B}efore model training {\bf (BIB)}, and {\bf B}ackdoor {\bf I}njection {\bf D}uring model updating {\bf (BID)}.
In the former setting, a new model is trained over the entire training dataset  before deployment. Hence, a small number of backdoor samples have to be injected into the original training dataset prior to model training. This setting could be applied to a variety of possible attacks scenarios, including  a malicious cloud handling outsourced training tasks or insiders/intruders who carry out injection into trusted data sources in a stealthy manner.
In addition, attackers are also motivated to train a model with a poisoned dataset and release it to potential victim users.

In a typical BID setting, a pre-trained model $M_{pre}$, as the victim model, already exists, and gets updated with data containing new information (backdoor samples in our case). 
The attacker can either pollute a publicly available pre-trained model offline, or inject crafted samples into newly collected data used to perform online training. For either BID case, backdoor samples are inserted into data batches in a sequential order, and used to update all the parameters of the CNN model.~\footnote{Mini-batch gradient descent is a common approach to train and update a CNN.
Particularly, for the latter case,
the attacker can initiate the injection procedure a few steps before he intends to use the backdoor, and keep injecting backdoor samples until the backdoor is no longer  needed.
The adversary can  therefore adjust the injection rate, amount and timing of injection as necessary. Once the injection stops, the backdoor is gradually removed  as the model continues to get updated with pristine data.}

\subsubsection{Poisoned Training}
Typically, a CNN model is trained and updated using mini-batch gradient descent optimization \cite{cotter2011better, bottou2012stochastic} and backpropagation \cite{lecun1998efficient}. The training procedure is essentially a process of finding the optimal weights of a neural network such that an objective function, commonly the cross entropy loss \cite{Goodfellow2016} between the ground truth label and prediction output in classification problems, is minimized. 
To enable a successful backdoor injection attack,   the training process of the model  is poisoned with \textit{backdoor} samples. Accordingly, the objective of the model training procedure  maximizes the accuracy of the training set and also the attack success rate of backdoor samples being ``misclassified'' as the target label.  This can be mathematically formulated as follows, assuming there is only one type of backdoor:
\begin{align}
\{W\} = \argmaxA_{\{W\}}\sum_{i\in D_{T}}\sum_{j \in N_C }y_{ij}*\log(Prob(pred=j|x_{i},\{W\})) \nonumber \\ + \sum_{i\in D_{A}}\log(Prob(pred=t|x_{i}^{b}+v,\{W\}))
\end{align}
where $x_i$ is the $i$th image in the corresponding dataset, $y_{ij}$ is the indicator of $x_i$ belonging to class $j$, $Prob(pred=j|x_{i},\{W\})$ is the model's output probability for class $j$ conditioned on the current parameter set $\{W\}$ and input $x_i$. The definition of the remaining notations are presented in Table \ref{tab:notation}.
We seek to evaluate the effect of poisoning under different scenarios in terms of adversary's knowledge and capability. The attacker is assumed to either have access to the original training data or to use a separate non-overlapping dataset.

\section{Backdoor Generating Strategies}
\label{sec:backdoor}

A perturbation, used  as a   backdoor, is the key to the success of the proposed backdoor injection attack. On the one hand, it plays an essential role in determining how effective the injection of backdoor samples is.  Ideally,  the backdoor pattern and its target class should  be easily learned and effectively recognized by the model after training. On the other hand, it is equally important that the perturbation mask is able to evade detection. 
The pixel intensity change introduced by the perturbation mask  should be as minimal as possible so that human eyes cannot differentiate between the original image and the perturbed one.

\begin{figure}[h!]
    \begin{subfigure}
        \centering
        \includegraphics[width=0.49\columnwidth]{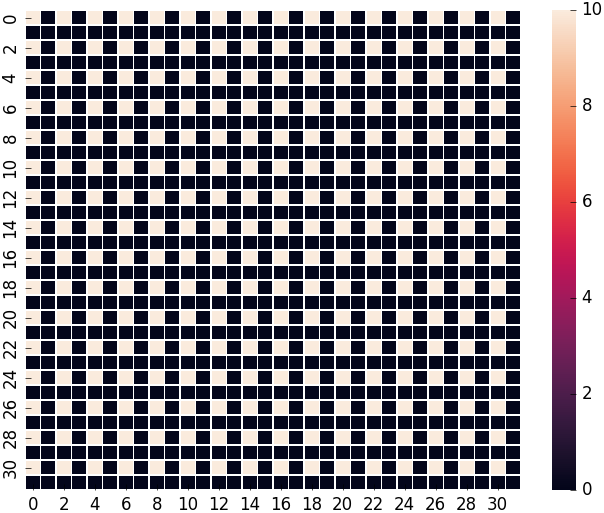}
    \end{subfigure}
    ~ 
    \begin{subfigure}
        \centering
        \includegraphics[width=0.49\columnwidth]{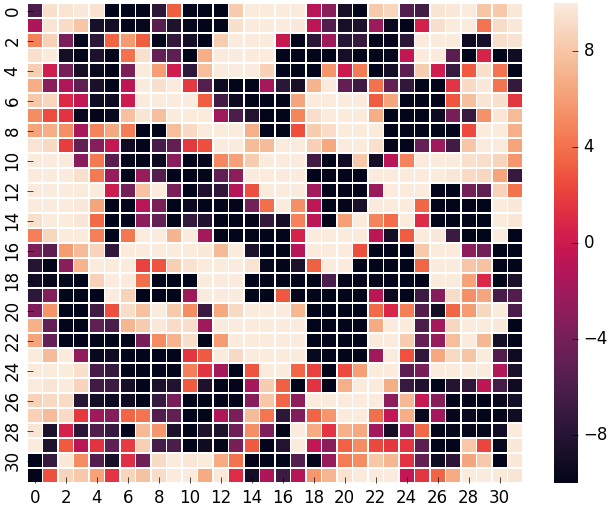}
    \end{subfigure}
    \caption{Examples of heatmap of the two types of perturbation masks. Left: patterned static perturbation mask where the intensity value increases by 10 at the position $(i_p,j_p) = (0,0)$ (i.e., the top-left light-pink pixel) within each $2\times2$ sub-region. Note that all the black pixels in the mask indicate no intensity change to an original image in the corresponding positions. 
    Right: adaptive perturbation mask with intensity change in both positive and negative directions, both with max intensity change $c_m = 10$.}
    \label{fig:perturbations}
\end{figure}

In this regard, we present two alternative approaches to develop a perturbation mask as the backdoor. Our first perturbation mask is one with a simple pattern built upon empirical observations. The second type of perturbation mask is generated based on a principled approach that systematically perturbs samples with small or even minimal intensity changes. We show the heatmap of the two types of perturbation masks in Figure \ref{fig:perturbations} as illustrative examples.

\subsection{Patterned Static Perturbation Mask}
\label{sec:static_pert}
The first type of backdoor we consider is a static perturbation mask generated based on a na\"{\i}ve approach. The intuition here is that  CNN models generally are able to learn imagery pattern features effectively,  as Convolutional Neural Network's  filters can quickly learn to exploit the strong spatially local correlations present in the images.
Accordingly, we can leverage a \textit{patterned static perturbation}, which can be treated as a new image pattern, as the backdoor for the CNN model to learn.

The patterned static perturbation mask works as follows. Firstly, given an   image $x$ of size $w \times h$, we generate a zero-value perturbation mask of equal size into multiple non-overlapping sub-regions of the same size $r \times r$ adjacent to each other. Then, within the first sub-region, one particular position $(i_p, j_p), 0 \leq i_p < r, 0 \leq j_p < r$, is randomly chosen, where we assign a constant value of intensity change, denoted by $c_m$. We apply the same value of intensity change to the same position in the next adjacent sub-region, and repeatedly do so for the remaining sub-regions.
As a result, we yield the static perturbation mask $v$ as:
\begin{equation}
  v_{ij}=\begin{cases}
    c_m, & \quad \text{if} \quad (i+i_p) \bmod r = 0, \quad (j+j_p) \bmod r = 0.\\
    0, & \quad \text{otherwise}.
  \end{cases}
\end{equation}
where $v_{ij}$ denotes the value of intensity change at $i$th row and $j$th column of perturbation mask $v$.  Note that a perturbation is introduced at only the single chosen position, in each sub-region.

\begin{figure}[h!]
    \centering
    \includegraphics[width=0.8\columnwidth]{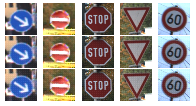}
    \caption{Examples of Images with Patterned Static Perturbation. (First row: original images. Second row: images with static perturbation mask of max intensity change $c_m = 6$. Third row: images with static perturbation mask of max intensity change $c_m = 10$.)}
    \label{fig:empirical_perturbation_example}
\end{figure}

To apply the perturbation, we simply add the perturbation mask $v$ to the original image $x$ to create a backdoor sample $x^b$, i.e. $x^b = x + v$.
To achieve stealthiness, the choice of value of intensity change should minimally perturb the original image, yet be strong enough for the model to learn the backdoor effectively. Empirically, we set different intensity values in our experiments as discussed in Section \ref{sec:evaluate}.
Here, we show some examples with static perturbation mask with intensity change equal to $6$ and $10$, respectively, in Figure \ref{fig:empirical_perturbation_example}.

\subsection{Targeted Adaptive Perturbation Mask}
\label{sec:adaptive_pert}

One limitation of the patterned static perturbation mask  is that it is based on   a  repeated pattern, regardless of content and classification models. As such,  the static perturbation mask may  not be optimal backdoor for the model to learn.

To improve, we devise a  second type of backdoor,  applied through an \textit{adaptive perturbation} mask,  that instead takes both the data and an existing model into consideration when generating the perturbation. The hypothesis is that  if the attacker is able to leverage this information, he can create a stronger  backdoor specific to the attack scenario at hand.

An intuition behind this approach is the following. As observed in recent studies \cite{moosavi2017universal}   
deep learning models generate regions of decision  boundaries that are nonlinear and that can be compromised by universal perturbations. 
Accordingly, we hypothesize that if we can find an adaptive perturbation that can ``push'' all the data points from  a given class    toward  the decision boundary of the  target class,   an  attack will have high chances of success, even with a small perturbation to the original image.



We implement our targeted adaptive perturbation according to this intuition, extending Moosavi's work \cite{moosavi2017universal}  on  CNN robustness against adversarial perturbations. Moosavi-Dezfooli et. al. proposed a novel approach  to compute a {\it single} adversarial perturbation that can be universally applied to random images from a given domain, inducing random (non-targeted) misclassifications on most
images from the domain. This concept of adversarial perturbations makes it a seemingly ideal candidate as the backdoor perturbation in our case. 
We have a substantially  different goal than \cite{moosavi2017universal}, however, seeking to achieve targeted misclassifications, and only on a niche set of images, and therefore need to apply a different approach   
\footnote{To achieve targeted misclassification, we can directly rely on our proposed algorithm to generate a perturbation as backdoor without injection in the training data, if we do not consider stealthiness (by setting a very large constraint). We will report the results in Appendix \ref{app:direct}}. 
Accordingly, we  propose a customized way to generate an adaptive perturbation mask,   explained in the following.

Let $X$ be the set of all data points in the given class $c$ from a training set, and let $f$ denote the classification function of a neural network model. The algorithm runs in an iterative fashion over all of $X$. At each iteration, it goes through each image $x_i$ in $X$, and computes the minimum perturbation {\it change} $\Delta v_{i}$ that pushes the image $x_{i}$ embedded with the current perturbation $v$ toward the decision boundary of target class $t$. In other words, if $x_{i} + v$ cannot enable targeted misclassification, we compute an extra perturbation $\Delta v_{i}$ by solving the optimization problem:
\begin{align}\label{opt}
    \Delta v_{i} = \argminA_r||r||_{2}, \quad s.t. \quad f(x_{i}+v+r) = t
\end{align}
Specifically, following  the derivation introduced in the Deepfool algorithm \cite{moosavi2016deepfool}, at each iteration we compute the minimum perturbation  matrix  $[\Delta v_{i}]$ that
projects the current data point to the boundary of the approximated polyhedron edge between the given class c and the target class $t$. The details are given in Algorithm \ref{alg2}.

\begin{algorithm}[h!]
 \caption{Targeted DeepFool algorithm}
 \label{alg2}
\KwIn{Data point $x$ from class $c$, classifier $f$, target class $t$, threshold of max iteration $I$}
\KwOut{Adaptive perturbation $v$}
\SetAlgoLined
Initialize $x_{0} = x$\;
$i = 0$\;
\While{$i \leq I$}{
    $w = \bigtriangledown f_{t}(x_{i}) - \bigtriangledown f_{c}(x_{i})$\\
    
    $f = f_{t}(x_{i}) - f_{c}(x_{i})$\\
    
    $v_{i} = \frac{|f|}{|w|_{2}^2}$\\
    
    $x_{i+1} = x_{i}+v_{i}$\\
    
    $v = v + v_{i}$\\
    
    $i = i + 1$

}
\end{algorithm}

To be stealthy, the magnitude of the adaptive perturbation $v$ should be constrained, i.e., $\|v\|_{\infty} \leq \xi$.
That is, we project the updated perturbation on the  $l_{\infty}$ ball of radius $\xi$    centered at 0. The projection function is defined as $ P_{p,\xi}(v) = \argminA_{v'}||v-v'||_{2}, s.t. \|v'\|_{\infty} \leq \xi$. Solving this problem will result in a perturbation with the max value of $\xi$.

After reaching the  max iteration threshold (we used the default value from \cite{moosavi2016deepfool}),
the process stops and yields our adaptive perturbation. Note that we do not need to generate an adaptive perturbation that directly renders a perturbed image $x_i + v$ from class $c$ to be misclassfied as target class $t$. Rather, it is sufficient to push
$x_i + v$ toward (or close to) the decision boundary of the target class -- it is anticipated that
the subsequent model learning, using these poisoned samples with their target class, will induce the desired misclassifications.
Thus, the amount of perturbation is limited: by setting a low magnitude constraint $\xi$, we generate an adaptive perturbation that is small enough to  be effectively learned by the victim model via data poisoning. The pseudo code of the algorithm is shown in Algorithm \ref{alg1}.

\begin{algorithm}[h!]
 \caption{Compute an adaptive perturbation for class $t$}
 \label{alg1}
\KwIn{Data points $X$ from class $c$, classifier $f$, desired $l_{p}$ norm of the perturbation $\xi$, target label $t$, threshold of max iteration $I$}
\KwOut{adaptive perturbation v}
\SetAlgoLined
Initialize $v = 0$\;
$i = 0$\;
\While{$i \leq I$}{
    $i = i + 1$\\
    \For{each data point $x_{i} \in X$}{
        \If{$f(x_{i}+v) \neq t$}{
            Compute the minimal perturbation that sends $x_{i}+v$ to decision boundary:\\
             Using Algorithm \ref{alg2}: $\quad\quad \Delta v_{i} = \argminA_r||r||_{2}, \quad s.t. \quad f(x_{i}+v+r) = t$\\
            Update the perturbation:\\
            $\quad\quad v = P_{p,\xi}(v+\Delta v_{i})$
        }
    }
}
\end{algorithm}

\begin{figure}[h!]
    \centering
    \includegraphics[width=0.8\columnwidth]{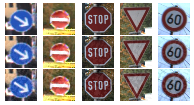}
    \caption{Examples of Images with Targeted Adaptive Perturbation. (First row: original images. Second row: images with adaptive perturbation of max intensity change $c_m = 6$. Third row: images with adaptive perturbation of max intensity change $c_m = 10$.)}
    \label{fig:perturbaiton_perturbation_example}
\end{figure}

Similar to the case of the  patterned  static perturbation, we directly add the perturbation to the original image to produce a backdoor sample. We show various examples of backdoor samples with perturbation generated from different settings in Figure \ref{fig:perturbaiton_perturbation_example}.

\section{Experimental Evaluation}
\label{sec:evaluate}
In this section, we describe the dataset, model architecture, metrics used for evaluation, methodology and the corresponding experimental results.

\subsection{Dataset}
We use the following  datasets for our experiments.

\textbf{GTSRB} (German Traffic Sign). This dataset is made up of color images of German traffic signs from 43 different classes. It has $39,209$ images for training and $12,630$ images for testing. Particularly, we use the preprocessed version of the original raw dataset, where each image is resized to $32 \times 32$ with three RGB channels.
Additionally, we perform data augmentation by applying similarity transformation via random rotation, scaling or translation. As as result, we end up generating $5$ times more extra training data based on the original training dataset. We use the augmented GTSRB for the following experiments.

In addition, we validate our algorithm's generality using the following two datasets:

\textbf{MNIST} \cite{lecun1998gradient}. It consists of $28 \times 28$ grayscale handwritten digit images from $10$ classes, i.e., digit $\{0 \dots 9\}$. It has a training set of $55,000$ instances and a test set of $10,000$ instances.

\textbf{CIFAR-10} \cite{krizhevsky2009learning}. It has $60,000$ of size $32\times32$ color images in $10$ classes with $6,000$ images per class. There are $50,000$ training images and $10,000$ test images. 
The result of these two dataset will be reported on section \ref{sec:general}.

\subsection{Model and Deployment}

The victim model we use for the GTSRB dataset is based on the convolutional network architecture proposed in \cite{sermanet2011traffic}. The surrogate model, if necessary,  for the adversary to generate targeted adaptive perturbation is based on the convolutional network architecture proposed in \cite{lecun1998gradient}.
The details of model architecture and parameter setting are summarized in Appendix \ref{app:models}.

The CNN model architecture, training and testing process, as well as injection procedure are implemented in Python based on the Tensorflow framework. 

\subsection{Metrics}
To  evaluate the effectiveness of our proposed backdoor injection attack, we  rely on the following measures.
\begin{itemize}
\item{\bf Attack Success Rate}  represents the percentage of backdoor instances drawn from test set being classified as the target label. A high attack success rate indicates an effective strategy of backdoor injection attack.

\item{\bf Test Accuracy Loss}  refers to difference of the classification accuracy on the test set between the poisoned model and the unpolluted model. In BIB setting, unpolluted model refer to learning a new model without inject backdoor samples, while in BID setting, it is the pretrained model. The former is expected to perform as close as possible to the latter.

\item{\bf Perturbation Stealthiness} aims to evaluate the quality of the backdoor perturbation, and yet it is difficult to quantify. Although an image with a static or adaptive perturbation is likely visually imperceptible (see Figure \ref{fig:empirical_perturbation_example} and \ref{fig:perturbaiton_perturbation_example}), it is also preferable for the perturbation to evade machine detection. In this regard, we consider two quantitative measurements to evaluate the quality of the perturbations.
\end{itemize}

\subsection{Setup}
\subsubsection{Summary of Attack Scenarios}
For each injection setting, we provide a summary of the  attack scenarios to be simulated with respect to different levels of the adversary's knowledge and capability as described in Section \ref{sec:threat}. 

For {\em BIB} setting, to make the attack more challenging, we always assume the adversary has no prior knowledge of the original model. However, we still consider the chance of an adversary having access to original training data. Hence, we simulate the following two cases.
\begin{itemize}
    \item {\bf BIB-PKD:} Training Data Known \\ 
    With partial knowledge of original training data, samples can be directly drawn from training data $D_{T}$ to generate the injection set $D_A$ with either static or adaptive perturbation. 
    \item {\bf BIB-MK:} Training Data Unknown \\ 
    In contrast, a separate (surrogate) data subset, not used for training, is used by the adversary to produce the injection set for both types of perturbation.  This data may come from the same (training) database or from a different one.
\end{itemize}
For both cases described above, a surrogate model $M_{sg}$ is assumed to be used by the adversary in order to generate our proposed targeted adaptive perturbation.

For {\em BID} setting, due to the existence of a pre-trained model $M_{pre}$, we consider the possibilities  of an adversary having knowledge of the training data or/and pre-trained model. 
\begin{itemize}
    \item {\bf BID-FK:} Training Data Known and Pre-trained Model Known.
    \item {\bf BID-PKD:} Training Data Known and Pre-trained Model Unknown.
    \item {\bf BID-PKM:} Training Data Unknown and Pre-trained Model Known.
    \item {\bf BID-MK:} Training Data Unknown and Pre-trained Model Unknown.
\end{itemize}
Similarly, the adversary can leverage the knowledge of original training data or model if known. Otherwise, a surrogate dataset or model will be utilized. Note that, we mainly simulate online updating with GTSRB dataset whereas the same attack can also apply to offline updating (see Section \ref{sec:eva_cifar10}).

\subsubsection{Splitting Data} We use the augmented version of the GTSRB dataset, which consists of $247,884$ traffic sign images in total. It is split into four non-overlapping parts, i.e., 85.4\% as major set $D_{major}$ ($211,734$), 9.5\% as minor set $D_{minor}$($23,520$), and 5.1\% as model testing set $D_{test}$ ($12,630$). The major set and minor set have a ratio of $9:1$ roughly. 

In the BIB  setting, major and minor sets are combined together as the training set, i.e., $D_T = D_{major} \cup D_{minor}$. Besides, backdoor samples are also drawn from the combined set, i.e., $D_A \subset (D_{major} \cup D_{minor})$. In contrast, assuming the adversary has no access to the training data, only the  major set $D_{major}$ serves as the training set while minor set $D_{minor}$ is available to the adversary to produce injection set, i.e., $D_T = D_{major}, D_A \subset D_{minor}$. In order to emulate the BID setting, half of the training data is used to pre-train a model and the remaining half is reserved as the new coming data to update the pre-trained model. Table \ref{tab:data} summarizes these settings. Note that, in each attack scenario, only $80\%$ of the corresponding training data are actually used for model training and the remaining $20\%$ serve as the validation set.

\begin{table*}[t!]\footnotesize
\centering
\caption{Makeup of Training \& Injection Sets in Various Scenarios}
\label{tab:data}
\begin{tabular}{|l|l|l|}
\hline
Scenario & Makeup of Training Set $D_T$  & Makeup of  Injection Set $D_A$\\ \hline
 BIB-PKD & $D_T = D_{major} \cup D_{minor}$ & $D_A \subset (D_{major} \cup D_{minor})$ \\ \hline
 BIB-MK & $D_T = D_{major}$ & $D_A \subset D_{minor}$ \\ \hline
 BID-FK/PKD & $D_T = D_{major} \cup D_{minor} = \{\underbrace{\frac{1}{2}D_T}_\text{train}, \underbrace{D_T - \frac{1}{2}D_T}_\text{update}\}$ & $D_A \subset \frac{1}{2}D_T$ \\ \hline
 BID-MK/PKM & $D_T = D_{major} = \{\underbrace{\frac{1}{2}D_T}_\text{train}, \underbrace{D_T - \frac{1}{2}D_T}_\text{update}\}$ & $D_A \subset D_{minor}$ \\ \hline
\end{tabular}
\end{table*}

\subsubsection{Training} When the injection occurs before model training (BIB), the victim model $M_{vt}$ is trained using the poisoned training data. We use the model that yields the highest accuracy on validation set as the resulting model. 
On the other hand, when injection occurs while the model gets updated (BID), we first pre-train a model $M_{pre}$ using half of the training set. Then, $M_{pre}$, as the victim model, gets updated with the other half poisoned by the injection set. Specially, for BID setting, the two metrics are computed after $250$ incoming batches are reached. 
The  injection is stopped at a comparatively early stage, in order to emulate the case where the adversary would like to limit the  amount of injected content - and try to evade detection.  This is summarized as follows.

\begin{description}[align=left]
    \item [BIB:] $\{D_T = D_T \cup D_A\} \xrightarrow{train} M_{vt}$
    \item [BID:] $\frac{1}{2}D_T \xrightarrow{pre-train} M_{pre}$\\ $\{M_{pre}, (D_T - \frac{1}{2}D_T) + D_A\} \xrightarrow{update} M_{vt}$
\end{description}
To be noted, the details of $D_T$ and $D_A$ in different cases are presented in Table \ref{tab:data}.
In addition, to train and update the model, Adam optimizer\cite{kingma2014adam} is used for all experiments with an initial learning rate of $0.001$. The maximum training epoch is $20$.

\subsubsection{Generating a Targeted Adaptive Perturbation Mask} Unlike the static perturbation that can be generated independently from the data, creating an adaptive perturbation as backdoor for a particular class $c$ requires the help of a designated model, as well as samples belonging to the same class  drawn from data available to the adversary, according to the assumptions in each attack scenario. 
We report details on the setup for generating adaptive perturbation masks in Appendix \ref{app:adaptive}.

\subsubsection{Attack Target} We select $5$ pairs of labels $ \langle c, t \rangle$ out of the $43$ dataset classes as shown in Table \ref{tab:pairs}. We  include among these five both random pairs and intentionally selected pairs with both similar and contrasting targets in terms of shape and color. For each pair, backdoor samples are drawn from one class $c$ and assigned the label of target class $t$.
\begin{table}[h!]\footnotesize
\centering
\caption{Five Pairs of Classes $ \langle c, t \rangle$}
\label{tab:pairs}
\resizebox{\columnwidth}{!}{%
\begin{tabular}{@{}c|ccccc@{}}
\toprule
 Class Notation & \multicolumn{5}{c}{Class Name} \\ \midrule
$c$ & \begin{tabular}{@{}c@{}} Speed limit \\ (60 km/h) \end{tabular}  & Yield & Stop & No entry & Keep right \\ \midrule
$t$ & \begin{tabular}{@{}c@{}} Speed limit \\ (120 km/h) \end{tabular} & \begin{tabular}{@{}c@{}} Dangerous curve \\ to the right \end{tabular} & \begin{tabular}{@{}c@{}} Speed limit \\ (110 km/h) \end{tabular} & Ahead only & Keep left \\ \bottomrule
\end{tabular}%
}
\end{table}

We consider  one pair at a time in our experiments. Hence, for each scenario, we repeat the same attack for each of the $5$ pairs and record their corresponding results of attack success rate and model accuracy loss with respect to the test set $D_{test}$.

\subsubsection{Injection Strategy} In the  BIB setting, we create a small number of backdoor samples and inject them into pristine training data at once (injection ratio varies from 1.7\% to 4.7\%). For the BID setting, only a handful of backdoor samples (injection number varies from 4 to 10) are injected into each batch (size of 128) of the incoming data in sequential order. 

\subsection{Evaluation of Backdoor Injection Attack Under Various Scenarios}
In this section, we present our evaluation of backdoor injection attack  for both types of backdoor perturbation. 
Given the setup described in the prior section, we conduct BIB and BID with  both strategies (see Section \ref{sec:backdoor}) with a fixed max intensity of $10$, while varying the injection number. For each perturbation and its corresponding  case, we evaluate the attack with the chosen $5$ pairs of classes. For each pair, we calculate the corresponding model accuracy loss and attack success rate based on the test set. 
Final results are averaged among the $5$ pairs and reported in Table \ref{tab:off_pert_emp} and \ref{tab:on_pert_emp} respectively.

\subsubsection{Attack Performance}
According to Table \ref{tab:off_pert_emp} and \ref{tab:on_pert_emp}, given the same scenario in either BIB or BID setting, adaptive perturbation (above $90\%$ mostly) generally outperforms static perturbation (below $90\%$) in terms of attack success rate.
When it comes to test accuracy loss, it is very small and consistently below or near $1\%$ if adaptive perturbation mask is used.
For the static perturbation mask, the accuracy loss on average in each setting is comparatively larger with a highest  loss around $3.1\%$, and attack is even not success in some scenario.  This partially validates our hypothesis that image masks crafted with adaptive perturbation are easier for the model to accommodate because of the construction of the perturbation, which accounts for the data and the model at hand.

Note  that for both types of perturbations, the injection attack in the BID setting has a greater impact on test accuracy than that in the BIB setting. This is  expected since  incremental learning of new data may negatively influence the previously learned information (catastrophic forgetting \cite{french1999catastrophic,mccloskey1989catastrophic}). Clearly,  attacks in  BIB setting  have no such problem since the crafted samples are  jointly learned with the original data. However, such impact (in the BID case) in general is still considerably limited, especially for adaptive perturbations.
Furthermore, the attacks in BID setting achieve similar success rate to corresponding scenarios in BIB setting except for the one with static perturbation in {\em BIB-MK}.

Next we will discuss in details the impact of various factors in the following.

\begin{table*}[t!]\footnotesize
\centering
\caption{Average Attack Success Rate (\%) and Test Accuracy Loss (\%) in BIB setting with adaptive perturbation and static perturbation (Max Intensity Change = 10) w.r.t  Total Number of Injected Backdoor Samples}
\label{tab:off_pert_emp}
\begin{tabular}{l|l|l|l|l|l|}
\hline
\multicolumn{1}{|l|}{ {\bf Perturbation \& Scenario}} &\backslashbox{{\bf Metric}}{{\bf Injection }} &\makebox[3em]{10000} &\makebox[3em]{8000} &\makebox[3em]{6000}&\makebox[3em]{4000} \\ \hline
\multicolumn{1}{|l|}{\multirow{1}{*}{Adaptive Perturbation}} & Test Accuracy Loss & 0.41 & 0.48 &	0.28 &	0.41\\ \cline{2-6} 
\multicolumn{1}{|l|}{BIB-MK}                  & Attack Success Rate & {91.6} &	89.13 &	{90.61} &	88.13\\ \hline
\multicolumn{1}{|l|}{\multirow{1}{*}{Adaptive Perturbation}} & Test Accuracy Loss & 0.26 &	0.22 &	0.28 &	0.20 \\ \cline{2-6} 
\multicolumn{1}{|l|}{BIB-PKD}                  & Attack Success Rate & {97.64} & 97.15 & 96.58 &	94.84 \\ \hline
\multicolumn{1}{|l|}{\multirow{1}{*}{Static Perturbation}} & Test Accuracy Loss & 0.36 & 0.83 & 0.48&	0.84\\ \cline{2-6} 
\multicolumn{1}{|l|}{BIB-MK}                  & Attack Success Rate &  50.85 &	54.48 &	48.02 &	22.02\\ \hline
\multicolumn{1}{|l|}{\multirow{1}{*}{Static Perturbation}} & Test Accuracy Loss & 0.82 &	1.0 & 0.62 &	0.64 \\ \cline{2-6} 
\multicolumn{1}{|l|}{BIB-PKD}                  & Attack Success Rate & {88.22} & 93.21 & 62.86 &	72.14 \\ \hline
\end{tabular}
\end{table*}


\begin{table*}[t!]\footnotesize
\centering
\caption{Average Attack Success Rate (\%) and Test Accuracy Loss (\%) in BID setting with adaptive perturbation and static perturbation (Max Intensity Change = 10) w.r.t  Number of Injected Backdoor Samples per Batch}
\label{tab:on_pert_emp}
\begin{tabular}{l|l|l|l|l|l|}
\hline
\multicolumn{1}{|l|}{\bf Perturbation \& Scenario} &\backslashbox{\bf Metric}{\bf Injection} &\makebox[3em]{10} &\makebox[3em]{8} &\makebox[3em]{6}&\makebox[3em]{4} \\ \hline
\multicolumn{1}{|l|}{\multirow{1}{*}{Adaptive Perturbation}} & Test Accuracy Loss & 1.16& 0.85 & 0.67 &  0.66\\ \cline{2-6} 
\multicolumn{1}{|l|}{BID-MK}                  & Attack Success Rate & {93.12} & {92.12} & {90.03} & {85.05} \\ \hline
\multicolumn{1}{|l|}{\multirow{1}{*}{Adaptive Perturbation}} & Test Accuracy Loss & 0.87 &	0.87 &	0.61&	0.62  \\ \cline{2-6} 
\multicolumn{1}{|l|}{BID-PKM}                  & Attack Success Rate &  {95.19} &	{93.96} &	{93.33}&	{84.88}  \\ \hline
\multicolumn{1}{|l|}{\multirow{1}{*}{Adaptive Perturbation}} & Test Accuracy Loss &  0.72 &	0.68 &	0.4&	0.5\\ \cline{2-6} 
\multicolumn{1}{|l|}{BID-PKD}                  & Attack Success Rate & {91.52} &	{92.82} &	{87.17} & {86.36} \\ \hline
\multicolumn{1}{|l|}{\multirow{1}{*}{Adaptive Perturbation}} & Test Accuracy Loss & 0.35&	0.36 &	0.27&	0.32  \\ \cline{2-6} 
\multicolumn{1}{|l|}{BID-FK}                  & Attack Success Rate & {95.96}&	{ 96.04} &	{95.76} & {94.01} \\ \hline
\multicolumn{1}{|l|}{\multirow{1}{*}{Static Perturbation}} & Test Accuracy Loss& 3.1&	2.7&	1.97&	1.95\\ \cline{2-6} 
\multicolumn{1}{|l|}{BID-MK}                  & Attack Success Rate & {80.64} &	70.15 &	67.27 &	51.47 \\ \hline
\multicolumn{1}{|l|}{\multirow{1}{*}{Static Perturbation}} & Test Accuracy Loss &  3.15& 2.7& 2.5&	1.3\\ \cline{2-6} 
\multicolumn{1}{|l|}{BID-PKD}                  & Attack Success Rate & {87.73} & 81.07 & 64.18 & 50.67 \\ \hline
\end{tabular}
\end{table*}

\subsubsection{Effect of Injection Intensity}
As shown in Table \ref{tab:off_pert_emp} and \ref{tab:on_pert_emp}, when the number of injected samples increases, the attack success rate generally increases  for both settings. This performance improvement is less drastic for  targeted adaptive perturbation than   static perturbation. In other words, adaptive perturbation can be equally effective at a relatively low injection rate compared to the static perturbation.  
For instance, in BIB with the weakest assumption on the adversary knowledge ({\em BIB-MK}), we  achieve an average attack success rate above $90\%$ by only injecting $6,000$ backdoor samples (injection ratio is $2.8\%$) with adaptive perturbation of  max intensity change as 10. Similarly, in  {\em BID-MK} settings, we can also achieve a comparably decent attack performance with  an injection rate of  only $1.4\%$. In comparison, the best result for static perturbation is $88.22\%$ in {\em BIB-PKD} and $87.73\%$ in {\em BID-PKD}. These performance values are  still slightly under $90\%$ assuming the attacker has  knowledge of training data as well as having a higher injection ratio. 
 We also note that in general the BID settings required less injection samples than BIB settings.

\subsubsection{Effect of Source of Injection Data and Knowledge of Pre-trained Model}
In BIB settings, we found that the performance of an attack carried out using a static perturbation mask is greatly affected by the source of injection data. For instance, the attack success rate improves from $50.85\%$ ({\em BIB-MK}) to $88.22\%$ ({\em BIB-PKD}) if the adversary is assumed to have access to the training data. The reason is that it takes a variety of original images crafted with the same static perturbation mask for the model to learn such new pattern effectively instead of simply overfitting the backdoor samples.
Such increase is less striking for targeted  adaptive perturbation in those cases (from $91.6\%$ to $97.64\%$ attack success rate). Because the adaptive perturbation mask, generated via the original model
, is implicitly recognized by the model already. A set of less diverse backdoor samples is enough to trigger the victim model to learn such adaptive pattern. 
To some degree, this confirms our hypothesis that it is easier for the CNN model to learn an adaptive perturbation. While in the BID setting, such effect plays a less important role for both types of perturbation.The difference may be due to the fact that in BID setting, since the attack can succeed with a small number of batch training iterations, injection data drawn from a smaller set can easily satisfy the requirement of data diversity.

Moreover, in the BID setting, for adaptive perturbation, the knowledge of the pre-trained model does not significantly affect the attack success rate. As we can see from Table \ref{tab:on_pert_emp}, regardless of the knowledge of training data, the average attack success rate is comparatively close between {\em BID-MK} and {\em BID-PKM}, as well as {\em BID-PKD} and {\em BID-FK}. According to the finding in \cite{moosavi2017universal}, the computed perturbation can generalize well across different neural networks. Comparably, the adaptive perturbation generated from a surrogate model can still be effectively learned by the victim model.

\subsubsection{Effect of Max Intensity Change}
\label{sec:effect_intensity}
\begin{figure}[h!]
    \centering
    \begin{subfigure}
        \centering
        \includegraphics[scale = 0.45]{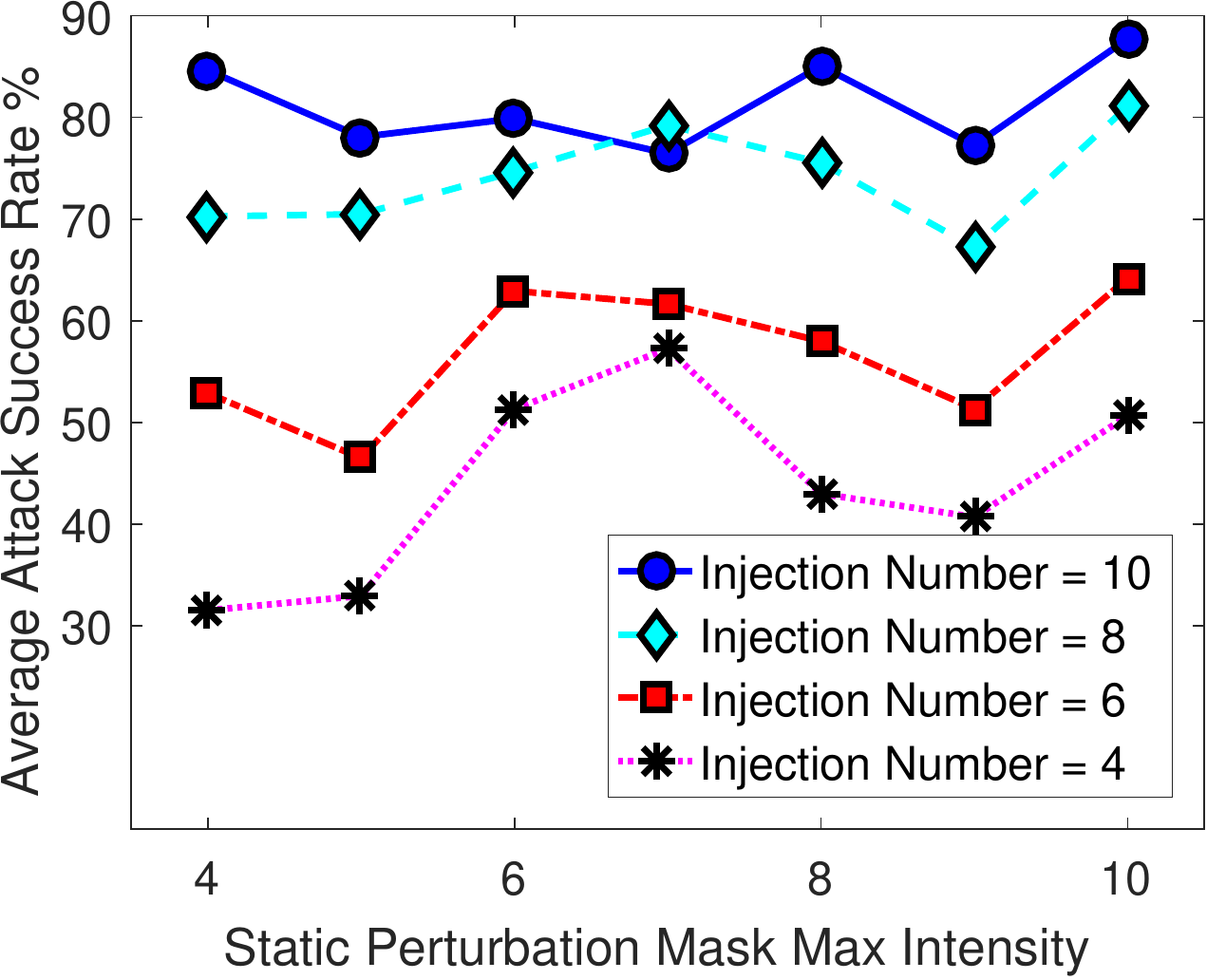}
    \end{subfigure}
    ~ 
    \begin{subfigure}
        \centering
        \includegraphics[scale = 0.45]{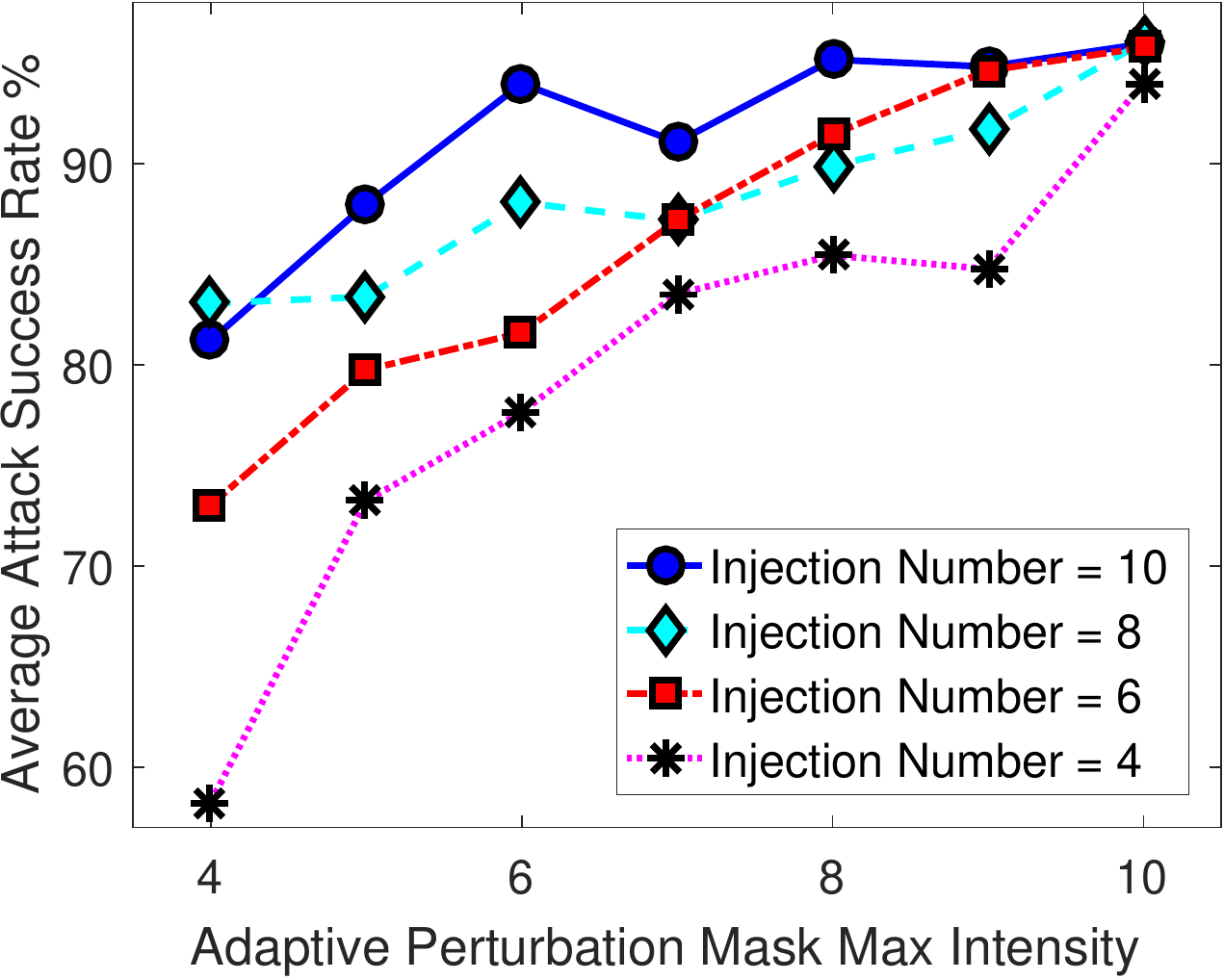}
    \end{subfigure}
    \caption{Effect of Max Intensity on Average Attack Success Rate with Static and Adaptive Perturbations}
    \label{fig:effect_intensity}
\end{figure}

In the previous experiments, we set a fixed value for max intensity of both perturbation approaches. Next, we  explore the effect of various perturbation max intensity values. Specifically, we choose a scenario {\em BID-FK} where adaptive perturbation achieves the best attack success rate. Comparably, we select {\em BID-PKD} for static perturbation with a similar setting and assumptions. For each case, in addition to varying the injection number from $4$ to $8$, we also set a perturbation max intensity change ranging from $4$ to $10$. 
We measure the average attack success rate as shown in Figure \ref{fig:effect_intensity}.

As we can see, for the  patterned static perturbation, the attack success rate fluctuates  for the same injection  rate. In contrast, for the targeted  adaptive perturbation, as the value of max intensity escalates, the attack success rate increases in a linear fashion. 

Targeted adaptive perturbation outperforms the static perturbation given the same setup of max intensity and injection number. Particularly, if  the injection number is factored in, adaptive perturbation with a larger max intensity can still be effective at a lower rate of injection. Given a high injection number (e.g., $8$ and $10$), the increase in max intensity only leads to a mild growth of attack success rate for the adaptive perturbation. But for static perturbation, the increased injection  rate significantly contributes to improving the attack success rate.

A possible explanation for these trends may be  that when creating a mask through adaptive perturbation,  a larger intensity change means the instance is much closer to the boundary of target class due to the adaptive nature of the pattern itself,  and therefore it is easier for the model to learn such adaptive pattern.  Yet, for the static mask, the intensity change does not necessarily push the instance in an optimal direction towards the target class's decision boundary since the patterned mask is generated independently regardless of characteristics of the images. Hence, a larger change might not contribute to rendering the static pattern easier for the model to learn. Besides, for both types of perturbation masks, it is straightforward that injecting more backdoor samples is helpful for the model to learn the corresponding backdoor. 


\vspace{2mm}
\noindent{\bf Summary of main findings} Based on the discussion in previous sections, we summarize some of the key findings in the following.
\begin{itemize}
    \item Generally, under the same conditions of max intensity and injection number, adaptive perturbation is more effective than static perturbation in producing a high attack success rate and low impact on model accuracy.
    \item On one hand, injecting more backdoor samples contributes to the success of the attack for both types of perturbation masks. On the other hand, it is worth noting that increasing the max intensity improves the attack efficacy of adaptive perturbation more evidently than static perturbation. Hence, by adjusting the two parameters properly one can yield a balance between achieving a good attack performance and preserving the stealthiness of the attack.
    \item The knowledge of training data has a greater impact on attack efficacy for the static perturbation in BIB setting than for BID setting. In contrast, the advantage of having access to the original model and training data is less striking for adaptive perturbation in both settings.
\end{itemize}

\subsection{Evaluation of Perturbation Stealthiness}
We  consider two popular metrics to evaluate the stealthiness of our perturbation approaches:
\begin{itemize}
\item
\textit{Perceptual Hashing (pHash) Similarity}: pHash \cite{pHash} represents a fingerprint of an image based on its features. Instead of focusing on the abrupt pixel change of an image, it reflects the overall feature representation. Images with similar features will have similar pHash value. We can calculate the similarity between the original image and perturbed image using the equation below to measure how much the original image is changed. \\
    $Similarity = \Big(1 - \frac{HammingDistance\big(PH_{ori}, PH_{new}\big)}{64}\Big) \times 100\%$\\
where $PH_{ori}$ is the pHash score of the original image, $PH_{new}$ is the pHash score of the backdoor image, and 64 represent the binary length of the pHash score.

\item
\textit{High Frequency Changes}: Fast Fourier transform converts the representation of an image from  the spatial domain to its frequency domain, where low frequency contains most basic information of an image while high frequency captures the significant changes. To measure the high frequency change, we first compute the Fourier transform of a given image or perturbation. Then we discard the low frequency part, and calculate the mean and standard deviation (stdev) of the L2 norm difference of the remaining part.
This is inspired by the approach proposed in \cite{jang2017objective}.
\end{itemize}

\begin{table}[h!]\footnotesize
\centering
\caption{A Summary of Metrics}
\label{tab:metric}
\resizebox{0.9\columnwidth}{!}{%
\begin{tabular}{@{}lccc@{}}
\toprule
& pHash & \multicolumn{2}{c}{High Frequency Change} \\ \cmidrule(lr){2-2} \cmidrule(lr){3-4}
& Similarity & Mean & Stdev \\ \midrule
Original Image & -- & $2.24\times10^7$ & $1.08\times10^8$ \\
Perturbed Image (Static) & $99.5\%$ & $2.25\times10^7$ & $1.07\times10^8$ \\
Perturbed Image (Adaptive) & $99.1\%$ & $2.23\times10^7$ & $1.06\times10^8$ \\ \midrule
Static Perturbation Mask & -- & $2.56\times10^4$ & $4.1\times10^5$ \\
Adaptive Perturbation Mask & -- & $2.59\times10^4$ & $4.9\times10^4$ \\ \bottomrule
\end{tabular}%
}
\end{table}

According to the results shown in Figure \ref{fig:effect_intensity}, the static perturbation with a max intensity change of $10$ has a similar performance as the adaptive perturbation with a max intensity change of $6$ at a fixed injection number of $10$ per batch. Thus, we compare those two perturbations generated for the $5$ pairs of $\langle c, t \rangle$ in Table \ref{tab:pairs}. For each pair, we select $100$ images belonging to class $c$ in the testing set to apply the corresponding perturbation, and measure the average pHash similarty, mean and stdev of high frequency distance between original and perturbed images. As to the two perturbation masks generated for a given pair, we only measure the mean and stdev of high frequency distance between them. The final results reported in Table \ref{tab:metric} are averaged among the $5$ pairs.

The results show that both perturbed images result in a very high pHash similarity score ($99.5\%$ and $99.1\%$) compared with the original one. Hence, the content of the original image is largely preserved. We also note that the image with static perturbation has a slightly  higher similarity value than that of the adaptive perturbation. This  is expected since the static perturbation changes only $1/4$ of all the pixels while adaptive perturbation is bound to  change a larger number of pixels. In addition, when it comes to the average mean and stdev of high frequency change,  the values for the perturbed image with both static and adaptive perturbation are extremely close to those of the original image. It again demonstrates that adding either perturbation does not significantly affect content of the original image.
Furthermore, if we look at the two perturbation masks directly, although they obtain a very similar mean value, the static perturbation has a much greater standard deviation of high frequency changes. It means the static perturbation changes certain image frequencies more drastically, indicating it is relatively less ideal compared to adaptive perturbation.

\subsection{Generalization of Backdoor Injection Attack}
\label{sec:general}
In this section, we further evaluate our proposed attack on two different datasets, i.e., MNIST and CIFAR-10, to see if our approach is still effective for other image classification tasks.

\subsubsection{Evaluation on MNIST dataset}
We begin our experiment with static perturbation using a simple CNN as the classifier (here we simply use the architecture introduced in Tensorflow tutorial\cite{tftutorial}). We start with the smallest value $1$ as its max intensity change to see if it is enough to create the backdoor for the model. For the target of attack, we select a pair of digit labels as $ \langle c, t \rangle = \langle 0, 2 \rangle$.

Interestingly, we only need to inject 10 backdoor samples to the training dataset applied with such a static perturbation in the BIB setting to reach nearly $100\%$ attack success rate without undermining the model accuracy at all. In BID setting, injecting 1 backdoor sample with such static perturbation in each batch is sufficient to yield $100\%$ attack success rate as well.

The results indicate the static perturbation alone is extremely effective to achieve a successful targeted attack. We believe this is due to the fact that images in MNIST generally have simple visual features and clean background, which can be easily affected by adding a tiny perturbation. Because of the enormous success of the static perturbation for MNIST, we do not try the adaptive perturbation. However, for images with very complex feature patterns such as those in GTSRB, a customized sophisticated adaptive perturbation can be more effective in rendering a targeted misclassification. 

\subsubsection{Evaluation on CIFAR-10 dataset}
\label{sec:eva_cifar10}
We adopt a state-of-the-art CNN model according to VGG-16 \cite{simonyan2014very}, which is further adapted to the CIFAR-10 dataset based on \cite{liu2015very}. This model is much deeper (and has far more parameters) than the model used for testing the MNIST dataset.
We consider the attack setting of injection during model updating (offline BID) where a public pre-trained model is provided and training data is also assumed known to the adversary. 

We use an injection  of $10$ backdoor samples per batch (size of $128$), and a max intensity change of $10$ for static and adaptive perturbation. We randomly choose a pair of class labels $ \langle c, t \rangle = \langle airplane, bird \rangle$ as the target of attack.  Our results show that both types of perturbation mask can achieve a success rate of over $98\%$ with an accuracy loss of $0.5\%$ after roughly $500$ batches.

In this experiment, we observed an interesting phenomenon during the training process. At first, the backdoor can be successfully learned by the CNN, but after a few rounds of batch updates with backdoor samples, the model accuracy drops significantly (roughly by $8\%$). After training with more batches, the model recovers its performance of accuracy and settles around $93\%$, while the attack success rate drops to $10\%$. As the training continues, the model maintains similar accuracy while increasing the attack success rate over $98\%$. We suspect that this may be due to the fact that with a deep structure of $16$ weight layers, the model is very sensitive to the original dataset at the beginning, and small weight changes affect performance notably. Yet, after being trained with more crafted samples along with the normal ones, the model can successfully recognize both of them, achieving good model accuracy and attack success rate simultaneously.

The above evaluations demonstrate that our proposed attack strategy generalizes well to different image classification tasks.


\section{Possible Defenses}
We  now discuss several defense strategies against our proposed attack. 
One straightforward idea to counter backdoor injection is to destroy the perturbation pattern. This can be accomplished according to several methods. For example, one may add some random noise to the test images, or may  blur the test images with a Gaussian filter. In our evaluation of these defenses, we consider the BID setting used in Section \ref{sec:effect_intensity}, and a fixed injection number of $10$ instances per batch. For both  perturbation approaches, we choose max intensity change of $10$ and $6$ respectively since they have similar attack performance according to Figure \ref{fig:effect_intensity}. Particularly, the noise added to every pixel has an intensity uniformly chosen from a range of $-20$ to $20$, and we use a $5 \times 5$ Gaussian filter implemented in OpenCV library \cite{opencv_library}. We note that both approaches decrease the test accuracy by (only) roughly $1\%$. With blurring, for static perturbation with max intensity change of $10$, the attack success rate  remains nearly unaffected. By contrast, for adaptive perturbation with max intensity change of $6$, it plummets by $34.6\%$. This indicates the non-robustness of adaptive perturbation under blurring. This is because blurring compromises the structure of the adaptive perturbation and induces a failure to push the data points over the decision boundary of a target class. An effective way to counter such defense is to use the blurred version of backdoor samples as the injection set, once the adversary is aware that blurring has been implemented. Notably, adding random noise has a limited effect on the attack success rate for both perturbations.

Another possible defense approach is to carry out a statistical analysis of the class frequencies of the training set given the fact that the attack requires injecting a number of backdoor samples associated with the same target label. However, this 
approach requires knowledge of the (true) class priors, and may also have only limited success when
the training set is not so large.
Additionally, in our experiment, we can carefully control the injection ratio to avoid injecting too many samples with the same label.  This may help to defeat such a defense. 

Another potential defense requires knowledge of ground-truth labels for the test samples (perhaps obtained by detailed human inspection and labeling).
Specifically, the
victim model will most likely generate a target class decision when an instance with the backdoor perturbation is presented in the testing phase. 
If human labels for these test instances are available (which will differ from the target class),
and if too many such ``misclassifications'' occur, the attack may be detected.
However, this approach is human-laborious as it requires test set labeling (whose avoidance is the
main purpose of using a classifier).  Moreover,
a crafty attacker may only infrequently exploit the backdoor. 
In such case the percentage of backdoor samples is likely to be extremely small with respect to a likely large volume of normal data. Hence, the target label does not necessarily prevail among other labels in the misclassified data.

A final possible defense could exploit high-dimensional clustering using deep layer feature information.
Specifically, if we consider the target class, the backdoor images may induce deep layer feature
patterns (in the trained deep network) that are very different from normal instances from the target class.  That is, in a deep
layer feature space, the target class may consist of two clusters, one associated with the backdoor and another associated with normal patterns.
Accurately identifying such clusters could help to identify the backdoor attack.  However, this
is a sophisticated, speculative defense strategy that is a good subject for future work.


\section{Related Work}
 We highlight some of the most relevant works in the area of adversarial machine learning  as follows.

{\em Evasion Attacks} are one of the well-studied attacks against machine learning models, where the adversary crafts adversarial samples that can fool the model at test time \cite{biggio2013evasion}. In the context of deep learning, Szegedy et al. \cite{szegedy2013intriguing} firstly noticed that applying an imperceptible perturbation to a test image can cause neural networks fail to classify it correctly. Subsequently, a number of studies \cite{goodfellow2014explaining, papernot2016limitations, moosavi2016deepfool, moosavi2017universal, jang2017objective} continue to refine the approach of generating {\em adversarial examples} to cause misclassifications given a model. \cite{papernot2017practical} even demonstrated such attack in a black-box manner. Aside from focusing on targeted misclassifications, in this work  we aim to create a backdoor that can be easily and universally applied, instead of customizing a unique perturbation for individual input instance.

Another line of research investigates {\em poisoning attacks}, whose aim is to poison the training data with malicious samples and degrade the efficacy of the resulting model. Poisoning attacks targeted traditional machine learning models \cite{biggio2012poisoning}, as well as deep learning models \cite{shen2016uror, munoz2017towards}. Our attack is a type of data poisoning {\it that can be exploited at test time}.  However,  our objective is
to embed the backdoor while {\it not} degrading the model's accuracy on regular data, unlike conventional data poisoning attacks. 
Also, in contrast to some poisoning methods that assume knowledge of the learning model or training data, our method can still work under weak assumptions where the attacker has quite limited
knowledge (some training examples). 

Some very recent works \cite{ji2017backdoor, Trojannn, liu2017neural} have  proposed a similar concept  as ours-- {\em neural network trojan attacks}. \cite{ji2017backdoor} directly manipulated the neural network parameters to create a backdoor.  By contrast, \cite{Trojannn} considers poisoning a publicly available model using training data generated via reverse engineering while \cite{liu2017neural} provides countermeasures against the trojan triggers in neural networks. Gu et al. \cite{gu2017badnets} study backdoor poisoning attacks in an outsourced training scenario where the adversary has full knowledge of the model and training data. Comparably, \cite{chen2017targeted} adopts a weak and realistic threat model assuming no knowledge of the training data and model. However, we notice that their generated trojan trigger or backdoor is not visually stealthy enough to not be detected, although various techniques (e.g., blending backdoor with original image \cite{chen2017targeted}, improving trigger transparency \cite{Trojannn}) have been applied.

Our work is also related to the field of image steganography \cite{johnson1998exploring, provos2003hide, cox2007digital, cheddad2010digital, shih2017digital} in the sense that we bear a similar goal of adding secret information or code to the image. However, the technique and application area is quite different from ours. In the future, it may be  worth exploring how to adopt the techniques in steganography for our attack purpose and see if deep learning model can learn such hidden information effectively.

\section{Conclusion}
In this paper, we propose a novel attack strategy against machine learning models named backdoor injection attack. Specifically, we design two kinds of stealthy perturbation masks as backdoors that can achieve high attack success rate with little influencing on the model's performance. Several realistic scenarios are
considered involving the threat model and when to inject the backdoor samples. Our detailed experiments demonstrate that our attack strategies are both stealthy and successful, and that  the choice of perturbation maximum intensity change and the injection rate affect to some limited extent the efficacy of our attacks. A potential refinement of our work includes injecting multiple different perturbation masks into a victim model at the same time, in order to make the attack harder to detect.  This could also allow for multiple backdoor targets.  We also would like to investigate applying variants of the proposed attack to other domains besides image classification.

\bibliographystyle{ACM-Reference-Format}
\bibliography{ref}

\appendix

\section{Setup for Generating a Targeted Adaptive Perturbation Mask}
\label{app:adaptive}
We list the setup of generating  adaptive perturbation masks (see Section \ref{sec:evaluate})  for the below cases given various assumptions on adversary's knowledge and capability.  $D_T$  varies by case  as  shown in Table \ref{tab:data}.
 \begin{description}[align=left]
     \item [BIB-PKD:] A surrogate model trained with $D_T$ is used. Samples of class $c$ are also drawn from $D_T$.
     \item [BIB-MK:] A surrogate model trained with $D_{minor}$ is used. Samples of class $c$ are also drawn from $D_{minor}$.
     \item [BID-FK:] The existing model pre-trained with $\frac{1}{2}D_{T}$ is used. Samples of class $c$ are also drawn from $\frac{1}{2}D_{T}$.
     \item [BID-PKD:] A surrogate model trained with $\frac{1}{2}D_{T}$ is used. Samples of class $c$ are also drawn from $\frac{1}{2}D_{T}$.
     \item [BID-PKM:] The existing model pre-trained with $\frac{1}{2}D_T$ is used. Samples of class $c$ are drawn from $D_{minor}$
     \item [BID-MK:] A surrogate model trained with $D_{minor}$ is used. Samples of class $c$ are also drawn from $D_{minor}$.
 \end{description}

\section{Model Structures and Training Setup}
\label{app:models}
We provide a summary of the model structures and parameters used. Notice, for the first three models, we use a dropout layer with keep probability of $0.5$ before the output layer.

\subsection{ConvNet for GTSRB}
This model is based on the architecture proposed in \cite{sermanet2011traffic}. It has $3$ convolutional layers and $1$ fully connected layers as shown in Table \ref{tab:convnet_traffic}. In particular, after the 3rd convolutional layer, its output is further concatenated with the output after the 2nd max pooling layer, which serve as the input of the last fully connected layer.

\begin{table}[h!]\footnotesize
\centering
\caption{Main Architecture of ConvNet for GTSRB}
\label{tab:convnet_traffic}
\scalebox{1}{%
\begin{tabular}{@{}cc@{}}
\toprule
Layer & Configuration \\ \midrule
1st Convolutional & filters=$6$, kernal size=$5\times5$, stride=$1$, activation=${\scriptstyle ReLu}$ \\
1st Max Pooling & kernal size=$2\times2$, stride=$2$ \\
2nd Convolutional & filters=$16$, kernel size=$5\times5$, stride=$1$, activation=${\scriptstyle ReLu}$ \\
2nd Max Pooling & kernal size=$2\times2$, stride=$2$ \\
3rd Convolutional & filters=$400$, kernal size=$5\times5$, stride=$1$, activation=${\scriptstyle ReLu}$ \\ 
Fully Connected & filters=$43$ \\ \bottomrule
\end{tabular}%
}
\end{table}

\subsection{LeNet-5 for GTSRB}
This model, served as the surrogate model, is based on LeNet-5 \cite{lecun1998gradient} with adaptation to GTSRB dataset. It has $3$ convolutional layers and $2$ fully connected layers as shown in Table \ref{tab:lenet_traffic}.

\begin{table}[h!]\footnotesize
\centering
\caption{Main Architecture of LeNet-5 for GTSRB}
\label{tab:lenet_traffic}
\scalebox{1}{%
\begin{tabular}{@{}cc@{}}
\toprule
Layer & Configuration \\ \midrule
1st Convolutional & filters=$6$, kernal size=$5\times5$, stride=$1$, activation=${\scriptstyle ReLu}$ \\
1st Max Pooling & kernal size=$2\times2$, stride=$2$ \\
2nd Convolutional & filters=$16$, kernel size=$5\times5$, stride=$1$, activation=${\scriptstyle ReLu}$ \\
2nd Max Pooling & kernal size=$2\times2$, stride=$2$ \\
3rd Convolutional & filters=$120$, kernal size=$5\times5$, stride=$1$, activation=${\scriptstyle ReLu}$ \\ 
Fully Connected & filters=$84$, activation=${\scriptstyle ReLu}$ \\
Fully Connected & filters=$43$ \\ \bottomrule
\end{tabular}%
}
\end{table}

\subsection{LeNet-5 for MNIST}
This model is based on LeNet-5 \cite{lecun1998gradient} as shown in Table \ref{tab:lenet_mnist}.

\begin{table}[h!]\footnotesize
\centering
\caption{Main Architecture of LeNet-5 for MNIST}
\label{tab:lenet_mnist}
\scalebox{1}{%
\begin{tabular}{@{}cc@{}}
\toprule
Layer & Configuration \\ \midrule
1st Convolutional & filters=$32$, kernal size=$5\times5$, stride=$1$, activation=${\scriptstyle ReLu}$ \\
1st Max Pooling & kernal size=$2\times2$, stride=$2$ \\
2nd Convolutional & filters=$64$, kernel size=$5\times5$, stride=$1$, activation=${\scriptstyle ReLu}$ \\
2nd Max Pooling & kernal size=$2\times2$, stride=$2$ \\
Fully Connected & filters=$1024$, activation=${\scriptstyle ReLu}$ \\
Output & filters=$10$ \\ \bottomrule
\end{tabular}%
}
\end{table}

\subsection{VGG-CIFAR10}
This model is based on VGG-16 \cite{simonyan2014very} with adaptation to CIFAR-10 dataset based on \cite{liu2015very}. It consists of 5 groups of convolution layers and 1 group of fully-connected layers with a total of 13 convolution layers and 2 fully-connected layers as shown in Table \ref{tab:vgg}. We use the same dropout configuration as \cite{liu2015very}. Besides, stochastic gradient descent (SGD) is used to optimize the model with an initial learning rate of $0.001$. We will stop updating the model once accuracy and attack success rate on validation set become steady (about 500 batches).

\begin{table}[h!]\footnotesize
\centering
\caption{Main Architecture of VGG-CIFAR10}
\label{tab:vgg}
\scalebox{1}{%
\begin{tabular}{@{}cc@{}}
\toprule
Layer & Configuration \\ \midrule
2 Convolutional & filters=$64$, kernal size=$3\times3$, stride=$1$, activation=${\scriptstyle ReLu}$ \\
Max Pooling & kernal size=$2\times2$, stride=$2$ \\
2 Convolutional & filters=$128$, kernel size=$3\times3$, stride=$1$, activation=${\scriptstyle ReLu}$ \\
Max Pooling & kernal size=$2\times2$, stride=$2$ \\
3 Convolutional & filters=$256$, kernal size=$3\times3$, stride=$1$, activation=${\scriptstyle ReLu}$ \\ 
Max Pooling & kernal size=$2\times2$, stride=$2$ \\
3 Convolutional & filters=$512$, kernal size=$3\times3$, stride=$1$, activation=${\scriptstyle ReLu}$ \\ 
Max Pooling & kernal size=$2\times2$, stride=$2$ \\
3 Convolutional & filters=$512$, kernal size=$3\times3$, stride=$1$, activation=${\scriptstyle ReLu}$ \\ 
Max Pooling & kernal size=$2\times2$, stride=$2$ \\
Fully Connected & filters=$512$, activation=${\scriptstyle ReLu}$ \\
Fully Connected & filters=$10$, activation=${\scriptstyle softmax}$ \\ \bottomrule
\end{tabular}%
}
\end{table}

\section{Generating Adaptive Perturbation mask For Direct Targeted Misclassification}
\label{app:direct}
As explained in Section \ref{sec:adaptive_pert}, the problem we tackle in this work is substantially different and difficult than that studied by \cite{moosavi2017universal} in the sense that we focus on targeted misclassification by means of a small perturbation and its magnitude must be well constrained for the sake of stealthiness. One may wonder why not solve the optimization problem \ref{opt} directly to achieve a targeted misclassification but instead generate an adaptive perturbation for poisoned training.
Here, we demonstrate that stealthiness cannot be guaranteed if targeted misclassification is enabled directly by the generated adaptive perturbation mask. Specifically, we choose $5$ pairs of targets $ \langle c, t \rangle$. For each of them, we vary the magnitude constraint from $10$ to $40$ to see if applying the generated adaptive perturbation mask to instances from class $c$ can cause misclassification as target class $t$ directly. The instances are drawn from the testing set and we measure the corresponding attack success rate . The results are averaged among the $5$ pairs and reported in Table \ref{tab:direct}.

\begin{table}[h!]\footnotesize
\centering
\caption{Attack Success Rate w.r.t. Various Magnitude Constraint on Adaptive Perturbation Mask}
\label{tab:direct}
\scalebox{1}{%
\begin{tabular}{@{}ccccc@{}}
\toprule
Magnitude Constraint & $10$ & $20$ & $30$ & $40$  \\ \midrule
Attack Success Rate & $1.6\%$ & $28.38\%$ & $74.68\%$ & $90.76\%$ \\ \bottomrule
\end{tabular}%
}
\end{table}

Accordingly, an adaptive perturbation mask with max intensity of $10$ can rarely succeed to cause targeted misclassification. In particular, to obtain an attack success rate above $90\%$, we have to apply a perturbation mask with max intensity as large as $40$, which would inevitably fail to achieve stealthiness considered as one of the most important design goals.


\end{document}